\documentstyle[aps, preprint, psfig]{revtex}
\def \hfb {\hfill\break}

\begin{document}
\draft
\title{\bf{The Bose Metal: gauge field fluctuations and scaling for field tuned quantum phase
transitions}}
\author{D. Das$^1$ and S. Doniach$^{1,2}$\\Departments of Applied Physics$^1$ and Physics$^2$\\
        Stanford University, Stanford, CA 94305. U.S.A.}
\date{\today}

\maketitle
\begin{abstract}
In this paper, we extend our previous discussion of the Bose metal to the field tuned
case.
 We point out that the recent observation
of the metallic state as an intermediate phase between the superconductor and the insulator in
the field tuned experiments on MoGe films is in perfect consistency with the Bose metal 
scenario. We establish a connection between general dissipation models and gauge
field fluctuations and apply this to a discussion of scaling across the quantum phase
boundaries of the Bose metallic state.  Interestingly, we find that the Bose metal
scenario implies a possible {\em two} parameter scaling for resistivity across the
Bose metal-insulator transition, which is remarkably consistent with the MoGe data.
Scaling at the superconductor-metal transition is also proposed, and a
phenomenolgical  model for the metallic state is discussed. The effective action of
the Bose metal state is described and its low energy excitation spectrum is found to
be $\omega \propto k^{3}$.
\end{abstract}
\newpage
\tighten

\section{Introduction}
In a recent paper\cite{dd}, we argued that a system of interacting Cooper 
pairs may form a gapless nonsuperfluid liquid, i.e. a metallic state, in two dimensions at
$T=0$. We called this state a Bose metal(BM). Although such an idea might seem
rather counter-intuitive offhand since a system of delocalised bosons
{\em typically} forms a superfluid(SF) state at $T=0$, closer thinking suggests
that it is not. A Bose metal is possible in the same sense that a spin liquid 
is possible in quantum antiferromagnets. In fact, we showed that it is just another
variety of a spin liquid: an $E_{2}$ spin liquid. 

In amorphous superconducting films, particularly MoGe, metal-like states analogous to the 
ones in granular superconductors\cite{dd}, have been observed in the presence of a magnetic field.
 The fact that such a  state exists as an intermediate 
phase between a superconductor(SC) and an insulator(INS), as predicted in Ref.\cite{dd}, is 
already a good hint
that the Bose metal scenario might be applicable to these systems as well.
In this paper, we generalize the arguments which were previously developed for the case of zero
applied magnetic field, to the field tuned case. We show that the results of field tuned
experiments in superconducting films, where such a metallic state has also been
seen\cite{nadya,ephron,valles}, also fit in with our concept of a Bose metal.

We first discuss (section II) the fact that gauge field fluctuations present in
the vortex  system in the JJA model
play a key role in the formation of the BM state and show how  generalized
dissipation models can also similarly lead to an analogous BM state. We then discuss how the concept
of the BM state may be generalized to the field tuned case. The observation of three
phases in ref.\cite{kap} is  consistent with the Bose metal scenario. This leads to
the suggestion of new scaling formulae for resistivity across the SC-BM and BM-INS
transitions. These are presented in section III and are shown to work  very well for
the MoGe data. In section IV, we develop a phenomenological model for the metallic
state based on the physics described above, which we find is in good agreement with
the MoGe data. Finally in section V we discuss the effective action and the low energy
excitation spectrum of the Bose metal state. Here, we also address the issue of why
the BM state has not been  observed in the previous analyses of the JJA model.

\section{The JJA model and effects of dissipation.}

\noindent
{\em Dynamical gauge field fluctuations}:
In ref.\cite{dd}, we considered a Josephson junction array (JJA) model
with onsite and nearest neighbour repulsion for large Cooper pair fillings
and zero external gate voltage (i.e. $\sum_{i}\left<\delta\hat{n}_{i}\right> 
= 0$, where $\delta\hat{n}_{i}$ is the charge fluctuation operator at the 
site $i$\cite{dd}). We showed using the duality transformation relations that this 
model maps onto a {\em two}-component {\em quantum} plasma
of vortices($V$) and antivortices($\bar{V}$), i.e. a set of 
{\em non-relativistic} bosons, moving
in a dynamically fluctuating gauge field $A^{\mu}$. 

This picture of vortices moving in a fluctuating gauge field is a
simple quantum mechanical extension of the results of classical phase fluctuations in a
2-d superconductor.
In the classical regime, a 2-d superconductor (with phase 
fluctuations) maps onto a two component classical plasma ($V\bar{V}$)
undergoing screening by a static electric field ($\vec{E} = -\vec{\nabla}
A_{0}$), as described by Kosterlitz and Thouless(KT)\cite{kt}. In the quantum regime
two new effects are important:\hfb
(a)the Bose statistics of the  vortices and antivortices allows for the
possibility of their superfluidity; and\hfb (b) because of quantum fluctuations, the
electric field is no more static  but dynamical, i.e. $\vec{E} = - \vec{\nabla}A_{0} -
(1/c_{s})\partial\vec{A} /\partial\tau$ and is associated with a corresponding magnetic field
$\vec{B} = \vec{\nabla}\times\vec{A}$.
Hence, expressed in terms of vortices, in the quantum
regime, two processes are important: depairing of $V\bar{V}$ pairs (or,
blowing up of vortex loops in (2+1)-d) and Bose condensation of depaired
vortices and antivortices, corresponding to the loss of phase order and
growth of charge order for the Cooper pairs, as discussed in ref.\cite{dd}.
The presence of dynamically fluctuating gauge field $\vec{A}$ causes  the
vortices and antivortices to pick up random Aharonov-Bohm(AB) phases $exp(i\int\vec{A}.d\vec{l})$
which makes these two processes distinct from one another. If there were
no fluctuating gauge field, then because of the quantum zero point motion, the
vortices and antivortices would Bose condense into a phase coherent state
as soon as they unbind, i.e. the film becomes insulating. Now because of the
gauge field $\vec{A}$, the situation here is different.
If the gauge field fluctuations are small, the dephasing induced by random AB
phases is weak and superfluidity of $V$ and $\bar{V}$ is retained. On the
other hand, if the gauge field fluctuations are very strong, then, because of the
random AB phases, the unbound $V$ and $\bar{V}$ fail to produce a phase coherent
state, and the system is a non-superfluid liquid.

The effects of these gauge field fluctuations may be seen very clearly in terms of a 
world line picture for the
vortices(appendix A). We represent this as a bosonic system interacting with a fluctuating gauge
field $\vec{A}$. The partition function is\cite{dd,rvb,paLee}:
\begin{eqnarray*}
   Z_{BA}&=&\frac{1}{N!}\sum_{P}\int_{\{r_{i}(\beta) = r_{Pi}(0)\}}
    \prod_{i}Dr_{i}(\tau) D\vec{A}(\vec{r},\tau)\exp\left(i\sum_{i}\int_{r_{i}(0)}^{
r_{i}(\beta)}\vec{A}.d\vec{r}_{i}\right)  \\
     &\times& \exp\left(-\int_{0}^{\beta}d\tau\left[
        \sum_{i}\frac{m}{2}\dot{r}_{i}^{2} + \frac{1}{2}\sum_{i\neq j}
v(r_{i} - r_{j})\right] \right) \exp\left(-S_{G}\left(\vec{A}\right)\right)
\end{eqnarray*}
where $S_{G}\left(\vec{A}\right)$ denotes the gauge part of the action and
$P$ is the permutation of the particles\cite{negele}.
This type of partition function may be obtained, for example, by
substituting $j^{0}(r,\tau)=
\sum_{i}q_{i}\delta(r - r_{i}(\tau))$ and $j^{\alpha}(r,\tau)=
\sum_{i}q_{i}\dot{r}_{i}^{\alpha}(\tau)\delta(r - r_{i}(\tau))$ in
eqn.(15) in ref.\cite{dd}, where $q_{i}$ is the vortex charge. The AB phase factor 
$\exp(i\int\vec{A}.d\vec{r}_{i})$ appears
explicitly here. When the gauge field fluctuations are weak, this factor
is close to $1$. So, this case is very similar to the Bose system without
any gauge field, and the ground state is an entangled liquid, i.e. a
superfluid (appendix A). Now, when the
gauge field fluctuations become very large,
the AB phase trapped by the bosons is of order $\pi$ (modulo $2\pi$), and the
phase factor
appears in $Z_{BA}$ with fluctuating signs. This would cause cancellation
of several terms in $Z_{BA}$ when entangled configurations are present,
implying that such configurations enter the partition function with a low
weight and are correspondingly high energy states, just like the
fermionic case (appendix A). So, in this case, the ground state is a disentangled
liquid, i.e. a non-superfluid.
Phase separation is not possible because of the long-range
interactions present in the (original) vortex Bose system. Hence, as gauge
field fluctuations
increases, there should be a phase transition from the superfluid to
a non-superfluid state.

 This discussion makes explicit why random AB phases causes disentanglement.
The non-superfluid state of bosons, thus obtained due to dephasing by random
AB phases, is a liquid rather
than a solid where the particles are localised by fluctuations, because
the gauge field fluctuates with time. Let us say that the vortices
get localised by fluctuations. Then, since they see a magnetic field that
fluctuates with time, by Faraday's law, they see an induced electric field
as well. This electric field will knock the localised particles out of their
positions. Thus, the presence of
a dynamically fluctuating gauge field in the quantum vortex system leads to
the possibility of a non-SF {\em liquid} phase, and by duality\cite{dhlee},
a consequent possibility
of a metallic state for the Cooper pairs, the original players in the system.
Because the vortices feel retardation effects due to gauge field fluctuations,
they dissipate
even in a pure system, thus leading to finite resitivity at $T=0$.

In summary, the effect of AB phases induced by the gauge field fluctuations is to
disorder the vortices and to create a {\em new} variety of quantum liquid, some 
signatures of which we discuss in this paper.

\noindent
{\em Effects of dissipation}:
The following four features about the Bose metal scenario emerge from the description
of the JJA model above and in Ref.\cite{dd}:\hfb
(i)when the Bose metal is observed, it exists as an intermediate phase between the SC 
and the INS;\hfb
(ii) the observation of the BM phase is associated
with two phase transitions: one from SC to BM and the other from BM to INS;\hfb 
(iii)the BM phase is dominated by dynamical gauge field(gf) fluctuations, as
felt by the vortices; and\hfb
(iv)there is a competition between gf fluctuations
and quantum zero point motion of the vortices at the BM-INS phase boundary: in
the INS phase, quantum zero point motion wins and the vortices are in the SF phase.

It has been suggested by Mason and
Kapitulnik\cite{nadya} that dissipation effects will quite generally help the formation
of a metallic like state for vortices in the quantum regime. We show here that  the
effects of a generalized dissipation model on the  quantum motion of the vortices
also maps onto a set of non-relativistic bosons moving in a dynamically fluctuating
gauge field, just like the JJA model.

Consider a generic dissipation model (no 
static disorder):
\begin{eqnarray}
 S = \int_{0}^{\beta} d\tau d^{2}r \frac{m}{2n}\mid\vec{j}(r)\mid^{2}&+&
\int_{0}^{\beta} d\tau d^{2}r d^{2}r'\rho(r,\tau)V(r-r')\rho(r',\tau)
\nonumber \\ &+& \frac{1}{2}\int_{0}^{\beta}
 d\tau d\tau' d^{2}r j^{\alpha}(r,\tau)\eta_{\alpha\beta}
(r-r',\tau-\tau')j^{\beta}(r,\tau')
\end{eqnarray}
where $j^{\alpha}(r,\tau) = \sum_{i}q_{i}\dot{r}_{i}^{\alpha}(\tau)\delta(r-
r_{i}(\tau))$, $\rho(r,\tau) = \sum_{i}q_{i}\delta(r-r_{i}(\tau))$ and
$n =$ average vortex density. $q_{i} =$ vortex charge, and the partition function
is $Z = \int Dr_{i}(\tau) e^{-S}$. The first
term is the quantum zero point motion term (also, known as the mass term);
the second term is the long-range (logarithmic, in a pure system)
interaction among the vortices and the last term is the dissipation term.
Summation
is implied in the dissipation term and $\alpha$, $\beta$ refer to the {\em space}
coordinates. This
action can be recast, by virtue of a Hubbard Stratanovich transformation on the
associated partition function, as:
\begin{eqnarray}
&S&=\int d\tau d^{2}r \frac{m}{2n}\mid\vec{j}(r)\mid^{2} +
\int d\tau d^{2}r d^{2}r'\rho(r,\tau)V(r-r')\rho(r',\tau)\nonumber \\ &+&
i\int d^{2}r d\tau j^{\alpha}(r,\tau)a^{\alpha}(r,\tau) +
\int d\tau d\tau' d^{2}r d^{2}r'a^{\alpha}(r,\tau) K_{\alpha\beta}
(r-r',\tau - \tau')a^{\beta}(r',\tau') \hspace{0.2in}
\end{eqnarray}
with $K_{\alpha\beta}(\omega_{n},k) = \eta^{-1}_{\alpha\beta}(\omega_{n},k)$
($\omega_{n}=$ Matsubara frequency). The dissipation kernel $K_{\alpha\beta}$
here is model dependent and the field $a^{\mu}$ is introduced through the
transformation. The Maxwellian type of coupling $\vec{j}.\vec{a}$ implies
that $a^{\mu}$ represents a gauge field. 
For example, for the Caldeira Leggett heat bath $K_{\alpha\beta}(\omega_{n},k) =
\delta_{\alpha\beta}\mid\omega_{n}\mid/\eta$, which is time-dependent. The dynamical nature of the
gauge field follows from the fact that dissipation is associated with 
velocity dependent forces. 

Clearly, this mapping implies two things. Firstly, as the vortices move in the 
presence of dissipation, they trap a random AB phase $exp(i\int\vec{a}.d\vec{l})$,
and consequently the arguments of dephasing by random AB phases presented earlier
in this section imply a disordered liquid phase for a dissipative system as well.
For example, if we consider the case of
Caldeirra-Leggett heat bath above, we see that: when dissipation $\eta$
is much weaker than the quantum zero point motion, gauge field flucutations
are weak, the randomness induced
by the AB phases is very small and the delocalised vortices are in an entangled state,
i.e. a superfluid. However, when dissipation is very strong,
gauge field fluctuations are also correspondingly very strong, and the
random AB phases can dephase the vortices into a disentangled, i.e.
non-superfluid, state. The liquidity of this state follows from the dynamical
nature of the gauge field. Because of this mapping, we shall
quite often use gauge field fluctuations and dissipation interchangeably
in this paper. Secondly, the above mapping means that a dissipative
quantum vortex system (where vortices have been induced by an external magnetic field)
has a very similar phase diagaram to that of the JJA model
considered in Ref.\cite{dd}, except that now dislocations are present (see below). There 
will be an SC phase which consists of dislocation - antidislocation pairs,
when quantum fluctuations are weak. As quantum fluctuations are increased
(by tuning the field, for example), the pairs will unbind. The state this film
would enter depends on the strength of dissipation. If dissipation is weak,
the vortices will form a superfluid state and the transition is from SC to INS.
On the other hand, when the dissipation is strong, the vortices would first enter
an uncondensed liquid like state due to dephasing effect by the AB phases.
By duality, this state is a non-SF liquid for the Cooper pairs (see first part of this section),
and hence, it is fair to call this a Bose metal. The resistance is induced 
by the free dislocations in this case. As the field is increased further,
quantum zero point motion overcomes dissipation (or, gf fluctuations), the
vortices would Bose condense and the film will be insulating. Thus, the transition
is SC-Metal-INS in this case. We explore
some of the consequences of this scenario below.

There are two key points which need to be kept track of
while generalising the results for vortices from the zero field case to the field 
driven case in a dirty film.
(a)In the presence of an external field, vortices enter an SC
film in the form of an Abrikiosov lattice. True long range order is not
possible in a 2d system at finite temperatures: dislocation-antidislocation
pairs are created in the Abrikosov lattice. As temperature is increased,
these pairs unbind (the analog of $V\bar{V}$ unbinding in the zero field
case), and (Cooper pair) superconductivity is destroyed.
(b)In the presence of static disorder, the Abrikosov
lattice is converted into a glass, called a vortex glass(VG). Energy
barriers between the various metastable states in this glassy phase are
{\em finite} in 2d\cite{fisher92,fisherB,geshk}. One way to understand this
is that dislocations are point like objects in 2-d, and that disorder
screens long-range log interaction between the dislocations. Thus, the
energy barriers to create dislocation-antidislocation pairs is finite in
2-d, and the energy barrier to their motion is very small, particularly in the
collective pinning regime\cite{geshk}, in which we mostly focus on here.
This means that as soon as the dislocation-antidislocation pairs are
created, they will hop around and induce finite resistance at any finite
temperature $\sim \exp(-\epsilon_{d}/kT)$, where $\epsilon_{d} =$ energy
barrier to creation of dislocation pairs. This would imply that
true superconductivity should set in at $T=0$. However, at $T=0$, tunneling
processes resulting from
quantum fluctuations due to quantum zero point motion, etc. can in
principle be strong and destroy long-range order in the vortex glass
phase, which is discussed in this paper.

\section {Scaling at the zero temperature phase transitions}
The presence of a Bose metallic phase in a SC 
film is associated with two phase transitions: one from SC to BM and another
from BM to INS. Correspondingly there will be two scaling behaviours for
the resistivity even for the field-tuned case, which we discuss below.

\noindent 1. {\em SC-BM transition}:
The first transition is associated with the unbinding of (quantum) dislocation - antidislocation
pairs (or, in a dirty system, when the {\em free} dislocation -
antidislocation pairs come into existence\cite{geshk}). The
film enters a metallic state due to strong gauge field fluctuations. 
Finite resistance in the film is induced by free dislocations, which
is proportional to the free dislocation density $n_{df}$\cite{geshk}:
$R_{\Box} \sim R_{Q}n_{df}\mu_{v}$. $R_{Q} = h/4e^{2}$ is the quantum 
of resistance and $\mu_{v} =$ vortex mobility. $n_{df}$ scales as 
$n_{df} \sim 1/\xi_{+}^{2}$. $\xi_{+} =$ SF correlation length that
diverges across the SC-BM phase boundary with an exponent $\nu_{0}$:
$\xi_{+} \sim (H - H_{c0})^{-\nu_{0}}$; $H_{c0}$ is the critical field
for SC-BM transition. Hence, on the metallic side,
\begin{equation}
  R_{\Box} \sim (H - H_{c0})^{2\nu_{0}}.
\end{equation}
This scaling formula is quite different from traditional quantum SC to
non-SC scaling $R_{\Box} = f(\delta/T^{1/\nu z})$. Observation of this scaling 
in ref.\cite{kap} provides good evidence that the metallic phase is a Bose 
metal. A comment about the value of $H_{c0}$: since energy barriers to 
metastable states are finite in 2-d in the presence of disorder, one might
think that a small amount of quantum fluctuations (zero point motion and
dynamical gauge field fluctuations) would destabilise the VG order
completely, i.e. $H_{c0} =0$. But, at very low fields, the vortex system
moves into the individual pinning regime from the collective pinning regime,
and the energy barriers to the vortex motion due to
pinning become very large\cite{geshk}. Hence, it is not inconceivable
that, in this very low field regime, at $T=0$, quantum fluctuations and
disorder can conspire to produce a pinned vortex state without any free
dislocations,
and consequently true (Cooper pair) superconductivity. In other words, there
is a possibility that $H_{c0} > 0$\cite{kap}.

\noindent 2. {\em BM-INS transition}: As the field is increased further,
quantum zero point motion of the vortices ($\sim \hbar^{2}n_{v}/m_{v}
\sim H$) increases, and beyond a critical value $H_{c}$, the zero point motion
overtakes the gauge field fluctuations, the vortices form a SF
phase and the film is insulating. Presence of free vortices during this
transition motivates a 
scaling for the resistivity across the BM-INS phase boundary. So far, this
phase boundary has been thought to be a SC-INS transition and the predicted
scaling\cite{fisherB} is $R_{\Box} = f(\delta/T^{1/\nu z})$. Although this
scaling formula works at high temperatures, it fails at low temperatures\cite{nadya}.
The {\em two} parameter scaling formula which we obtain below for the BM-INS
transition, however, scales the data across the entire temperature range, both high and low, when
the external field is in the critical regime.

In ref.\cite{dd}, we mentioned that
there is a jump in the vortex superfluid density at the BM-INS phase boundary.
Naively, this would imply that this is either a first order or a KT type
transition. But, the transition is possibly more subtle than this because it
is a phase transition from a (vortex) {\em superfluid to a gapless
nonsuperfluid phase}. In a first order transition, there is no divergent
length scale and the length scale is finite through the transition. But, the
BM phase is gapless, i.e. characterised by a diverged length scale.
The KT transition, which is a close precedent to this, on the other
hand, is a transition from a superfluid to a {\em gapped}
nonsuperfluid state. Also, the situation here is fundamentally different
from KT because of the involvement of the gauge degrees of freedom. Thus,
we suspect that this transition belongs to a very new  universality class. In
the absence of the knowledge of what exactly this universality class
is, it is hard to write down the scaling formula across this BM-INS
transition. Below, we suggest a way to scale the resistance across this metal
insulator transition based on phenomenological considerations and leave
the rest to future research.

Since this is a second order quantum phase transition, there is a diverging
correlation length $\xi$ with an exponent $\nu$ at this transition, i.e.
 $\xi \sim \mid \delta \mid^{-\nu}$, and a frequency scale
$\Omega$, which goes to zero with an exponent $z$, $\Omega\sim\xi^{-z}$,
where $\delta = (H - H_{c})$. As the energy dissipated scales as,
$  (V^{2}/R)t \sim \Omega$
($V=$ voltage drop, $t$ represents the time), resistance $R$ scales as
$    R\sim V^{2}/\Omega^{2}.  $
Since the dissipation is due to the vortices, the voltage induced by
moving vortices is
$  V = (h/2e)d\theta/dt,   $
with
$  d\theta/dt = 2\pi n_{f}Lv   $
\cite{halp}, where $v$ is the vortex velocity, $L$ is the length over which
the vortices move and $n_{f}$ is the vortex density. To obtain the scaling
of $V$, we note that the scaling of $L$ is
$   L\sim \xi \sim \Omega^{-1/z}.  $
Since $mv^{2}\sim \Omega$, the scaling of $v$ is
$   v\sim \sqrt{\Omega}.  $
We assume that the scaling of $n_{f}$ is
$ n_{f}\sim\mid\delta\mid^{\alpha}, $
where $\alpha$ is an exponent to be determined below. $n_{f}$ is {\em not}
the total vortex density $H/\Phi_{0}$, but rather the critical fraction
of the field induced vortices that participate in the dissipative process.
Combining all these factors, we find that the scaling of $R$ is
$  R \sim \delta^{2\alpha}/\Omega^{1+2/z}.  $
At any finite temperature, the divergence of $\xi$ is cutoff by temperature
$T$, i.e.
$  \Omega\sim T.     $
This also implies that the scaling function is always a function of
$\delta/T^{1/\nu z}$, i.e.,
\[   RT^{1+2/z}/\delta^{2\alpha} = f(\delta/T^{1/\nu z}) \]
where $f$ represents the scaling function. The right hand side comes from
the fact that $\xi \sim \delta^{-\nu}$, $\Omega \sim \xi^{-z}$ and
$\Omega \sim T$, so that $\delta/T^{1/\nu z}$ is the scaling variable.
To obtain $\alpha$, one needs
to first note that the resistance saturates to finite values independent of
temperature at low temperature; i.e. in the above eqn, we must have
$f(x) \rightarrow x^{-\nu(z+2)}$ in this limit.
Plus, since this low temperature resistance of the film
is {\em non-critical} through the metal insulator transition when $H$ is tuned
through $H_{c}$ as can be seen from eqn.(11) below, we should have
\begin{equation}
2\alpha = \nu (z + 2)
\end{equation}
Thus, the scaling formula for the resistance is:
\begin{equation}
 R\left[\frac{T^{1/\nu z}}{\delta}\right]^{\nu (z + 2)} =
     f(\delta/T^{1/\nu z})
\end{equation}
This is essentially a {\em two} parameter scaling formula, which is
expected of a bosonic system\cite{wlb}. For MoGe films\cite{nadya},
$z=1$ and $\nu = 4/3$, and hence, $2\alpha = 4$.
The corresponding plot of eqn.(5) for these films is shown in fig.1.
The data collapse with this two parameter scaling formula is quite
remarkable. We think that this points out that there is a true
metal-insulator transition at this critical field.
Although resistance usually does not receive scaling under normal 
circumstances, we found at least one precedent to this in the literature, i.e.
when a dangerously irrelevant variable is present\cite{dangerous}.

\section{ Phenomenology of field tuned experiments}
In this part, we discuss the phenomenology of the metallic
state in a dirty SC film, close to the metal-insulator transition, by setting up a
quantum ``pseudo-temperature" model\cite{ephron}. This is based on the idea of
competition between dissipation and quantum zero point motion of the 
vortices present near this transition point, presented so far.

The energy barrier to create
of dislocation pairs in a disordered vortex lattice is\cite{geshk}:
$  \epsilon_{d} \sim \epsilon_{0}ln(H_{c}/H).  $
At high temperatures, dislocations are created by thermal fluctuations,
i.e. dislocation density is
$ n_{dc} \sim \exp(-\epsilon_{d}/kT)$,
and the resistance induced by these is
$  R_{\Box} \sim R_{Q} n_{d} \mu_{v} \sim e^{-\epsilon_{d}/kT}.$
At low temperatures, quantum fluctuations take over, which we model in
terms of a quantum (pseudo) temperature $T_{Q}$: when $T_{Q}$ is small, the vortices
being bosons are in a SF state, and when high, it is in a disordered/metallic
state (real temperature remaining constant and small).
Because whether the vortices are in a SF state or non-SF state is
determined by the competition between the dissipation $\eta$ and quantum
zero point motion, $T_{Q}$ should be a function of these two parameters.
Since heating a system
increases its entropy, and (a)enhanced quantum zero point motion
($\sim n_{v}/m_{v} \sim H$) leads to superfluidity in a Bose system,
and hence, reduced entropy, while (b)action of dynamical gf fluctuations ($\sim \eta$)
increases (pseudo) entropy, we expect $T_{Q}$, for not too small fields, to scale as
\begin{equation}
T_{Q} \sim \frac{\eta^{\alpha}}{(n_{v}/m_{v})^{\beta}} \sim
\frac{\eta^{\alpha}}{H^{\beta}}
\end{equation}
where $\alpha, \beta$ are constants $>0$. (It can be any monotonic function of
$\eta/H$ in general.) Currently, we do not know how to obtain $\alpha$ and
$\beta$. In what follows, we shall assume $\alpha = \beta = 1$ and show
that this quantum temperature model captures the general phenomenlogical
features of the field-tuned expts in the aforementioned regime very
well.

Thus, expressed in terms of the quantum temperature model, as the field increases,
the quantum temperature of the system decreases and when it falls below a temperature
of the {\em order} of Kosterlitz Thouless temperature $T_{KT}$ (since vortices
here are 2-d bosons), the vortices would
form a SF, i.e. the film would be insulating. Since $T_{KT} \sim
\hbar^{2}n_{v}/m_{v} \sim H$,
the critical field $H_{c}$ above which the film is insulating is
obtained when $T_{Q} \sim T_{KT}$, i.e. $H_{c}$ scales as
(using eqn.(6)):
\begin{equation}
H_{c} \sim \sqrt{\eta} \sim 1/\sqrt{R_{n}}
\end{equation}
since $\eta$ scales inversely as $R_{n}$.
This makes sense physically, since as $\eta$ increases, gauge field
fluctuations increase and it is harder to create a SF out of the vortices.
This inverse dependence of $H_{c}$ on $R_{n}$ is broadly consistent with
the trend observed in the MoGe films (see Table I in Ref.\cite{yazdani}).

When $T_{Q} > T_{KT}$, the vortices are in a non-SF state and the film
displays a metallic response. From eqn.(6), the density of quantum
dislocations, by analogy with the thermal case, is then
$  n_{dQ} \sim \exp(-\epsilon_{d}/kT_{Q})  $.
Thus, the resistance of this metal at a finite temperature is given by
$R_{\Box} \sim R_{Q} (n_{dC} + n_{dQ})\mu_{v}$, i.e.
\begin{equation}
R_{\Box} \sim R_{n}[e^{-\epsilon_{d}/kT} + e^{-\epsilon_{d}/kT_{Q}}]
\end{equation}
This simple additive estimate is a good first hand measure when both
thermal activation and quantum tunneling processes are going on simultaneously
\cite{weiss}. Here we have used $\mu_{v}\sim R_{n}/R_{Q}$, which is
true of the Bardeen-Stephen processes. This implies that when the
real temperature $T$ falls below $T_{Q}$, the vortices enter the quantum
regime. This determines the crossover temperature $T_{cross}$ at which
the system moves from the classical to the quantum regime:
\begin{equation}
 T_{cross} \sim T_{Q} \sim \eta/H
\end{equation}
Hence, as the field is increased,  the crossover temperature decreases, as seen
in the Mo-Ge experiments\cite{ephron,nadya}. This unusual crossover from the
classical to quantum regime, which is very unlike that of a single
particle system, results from the strong cooperative
effects present in the vortex Bose system.

Thus, from eqn.(8), in the quantum regime, the resistance saturates to
\begin{equation}
R_{\Box} \sim R_{n} e^{-\epsilon_{d}/kT_{Q}} \sim
R_{n}exp(-C\frac{H}{H_{c2}}\frac{R_{n}}{R_{Q}}ln(\frac{H_{c}}{H}))
\end{equation}
where we have used dimensionless expressions for $\eta$ and $H$
by rescaling them with $R_{Q}$ and $H_{c2}$ respectively. $C$ is a
constant of order unity. Eqn.(10) has the same field dependence as that
obtained in Ref.\cite{shimsoni} if one expands the logarithm about the critical
field $H_{c}$, i.e. $ln(\frac{H_{c}}{H})\approx(\frac{H_{c}-H}{H})$,
\begin{equation}
R_{\Box}\sim R_{n}exp(-\tilde{C}\frac{R_{n}}{R_{Q}}(\frac{H_{c}-H}{H_{c}})),
\end{equation}
It should be noted
that the metallic resistance is {\em non-critical} across the metal-insulator
transition.

Thus, despite the simplicity of the quantum temperature model, it captures
the basic phenomenology of the experiments very well.

To summarise, we find several indications here, viz. (a)occurence of the metallic state as an 
intermediate phase between SC and INS phases, (b)scaling behaviour at the SC-Metal and
Metal-INS transitions being in accordance with the Bose metal scenario predictions and (c)the fact that the
phenomenology of the metallic state is as would be expected for a Bose metal state, to suspect that
the metallic state observed in the magnetic field tuned experiments in the SC films
is prrobably an adiabatic continuation of the Bose metal state presented in ref.\cite{dd}.

\section{The effective gauge field action of the Bose metal}
In this final section, we discuss the effective gauge action of the Bose metal based
on our analysis of the JJA model in Ref.\cite{dd}. We obtain certain properties
of this metal on the basis of this action, including its low energy excitation spectrum. And
then, we revisit the issue of why the Bose metallic
phase has not been observed in the previous analyses of the JJA model or
the Bose Hubbard (BH) model.

\noindent
{\em Effective action}: It is clear from the discussion in section II that a Bose metal is a liquid of
uncondensed bosons(vortices) moving in a transverse gauge field. The
effective action of this liquid can be obtained by integrating out the
vortices in a one-loop approximation. This calculation has already been
done in appendix E of ref.\cite{dd}: the transverse part of
the gauge action can be read off directly from eqn.(E.3) there. Thus,
the effective action of a BM is:
\begin{equation}
  S(\left[A^{\mu}\right]) = \sum_{\omega_{n}}\int d^{2}q\left[ \left(q^{2} +
  1/\xi_{+}^{2}\right)A^{0}_{\omega_{n},q}A^{0}_{-\omega_{n},-q} +
\left( \tilde{a} \frac{\mid\omega_{n}\mid}{q} + \tilde{b}q^{2} \right)
    \left( \delta_{\alpha\beta} - \frac{q_{\alpha}q_{\beta}}{q^{2}}\right)
    A^{\alpha}_{\omega_{n},q}A^{\beta}_{-\omega_{n},-q} \right]
\end{equation}
where the coefficients $\tilde{a}$ and $\tilde{b}$ are finite and scale with free vortex
density and may be read off from
eqn. (E.3) in \cite{dd} and $\omega_{n}=2\pi nT (n=$integer$)$ are Matsubara frequencies.
The spectrum of the longitudinal part of the gauge field
follows from the fact that since the vortices and antivortices are free,
they screen each other. Since the screening length scales as the density
of free vortices $n_{f}$ and $n_{f} \sim 1/\xi_{+}^{2}$, the above result
is obtained. Physically speaking, the above spectrum of the transverse gauge
field comes about,
beacuse it is dynamically screened\cite{tsvelik} and that there is no
spontaneous symmetry breaking in this phase. Thus, the longitudinal modes
are gapped but the transverse modes are gapless.

Since the gauge field $A^{\mu}$ is seen by the vortices,
one can use the duality transformation techniques\cite{dhlee,thesis} on this gauge action
to calculate the properties of the {\em Cooper pairs} (the original
bosons) in the Bose metal phase.
The superfluid density and the compressiblity of
the charges, i.e. Cooper pairs, are\cite{dhlee,thesis}:
\begin{eqnarray}
 \frac{1}{m}\rho_{s}^{c} &=& \lim_{q\rightarrow 0}\lim_{\omega\rightarrow
0} \frac{q^{2}}{q^{2} +1/\xi_{+}^{2}}  =  0  \\
  \kappa^{c} &=& \lim_{q\rightarrow 0}\lim_{\omega\rightarrow 0} \frac{q^{2}}{
  -i\tilde{a}\frac{\omega}{q} + \tilde{b}q^{2}} = \frac{1}{\bar{c}^{2}}
   = \textup{\hspace{0.0cm} (finite)}
\end{eqnarray}
Here the superscript $c$ denotes charges and we have used $\tilde{b} =
\bar{c}^{2}$ from eqn.(E.3), where $\bar{c}$ is the renormalised plasmon
velocity. These results are consistent with the fact that a Bose metal is a non-superfluid
compressible liquid.

The specific heat of the BM can be calculated from eqn.(12). This has already been
done in ref.\cite{dd}.
The gapless transverse modes make the most important contribution to the
specific heat: the low temperature specific heat of the BM goes as
$C \sim T^{2/3}$. This $T^{2/3}$
power law specific heat, in turn, directly implies that the low energy
excitation spectrum of the original Cooper pair system in this liquid
state goes as:
\begin{equation}
  \omega \propto k^{3}
\end{equation}
This unconventional excitation spectrum results from the strong correlation
effects present in the system and is consistent with the fact that
Landau critical velocity should be zero in a nonsuperfluid Bose system.
The density of states is correspondingly divergent: $N(\omega) \sim \omega^{-1/3}$.
Obviously, the BM state {\em cannot} be continued to a Fermi liquid or
mapped onto a set of free particles. We shall argue below that this liquid
is gapless beacuse it is an $E_{2}$ spin liquid, rather than an $SU(2)$
spin liquid.

In ref.\cite{dd}, we pointed out that the phase ($\phi_{i}$) and charge
fluctuation operators ($\delta n_{i}$) entering the JJA model are the
generators of $E_{2}$, the Euclidean group in two dimensions; $E_{2}$
being a group contraction of $SO(3)$. In this sense, the JJA model is
an $E_{2}$ ``spin" model and the BM phase being the disordered phase of
this model should be regarded as an $E_{2}$ spin liquid. An
extremely important point is that, whereas an $SU(2)$ spin liquid
is usually gapped, an $E_{2}$ spin liquid is {\em necessarily gapless}.
This follows from the distinction that whereas an $SU(2)$ spin model
maps onto bosons in a {\em constant} magnetic field\cite{kalmeyer1},
an $E_{2}$ spin
model maps onto bosons in a {\em fluctuating} magnetic field. Because of
the constant magnetic field, the $SU(2)$ spin liquid maps onto a Quantum
Hall liquid\cite{kalmeyer1}. This
means that the effective gauge action has a Chern Simmons term and the
excitations are correspondingly gapped. On the other hand, the effective
gauge action of the $E_{2}$ spin liquid is of Maxwellian type, i.e.
{\em without} any Chern Simmons term:
\[ S(\left[\vec{A}\right]) = \int d\omega d^{2}q f(\omega,q)A^{\alpha}_{
        \omega,q}A^{\alpha}_{-\omega,-q} \]
Now, if the excitation spectrum is gapped, i.e. $\lim_{\omega,q\rightarrow 0}
f(\omega,q) =$ constant, then this means that there is spontaneous
symmetry breaking and the z-component of the $E_{2}$ spins are ordered,
i.e. the system is charge-ordered\cite{thesis,dd}. Hence, the $E_{2}$ spin
liquid is necessarily gapless. For the JJA model we have been discussing
so far,
this feature can be seen directly by focussing on the transverse part of
the gauge action in eqn.(12).

It is hard to rule out the possibility of excitations with fractional quantum 
numbers\cite{ioffe1} in this liquid. However, since 
there is no explicit time reversal symmetry breaking in the effective action,
we think that such excitations would always exist in pairs being 
held together strongly by gauge-like forces.

\noindent
{\em JJA model: what's missing?}
In this last part of the paper, we revisit the issue of why the Bose metallic
phase has not been observed in the previous analyses of the JJA model or
the Bose Hubbard (BH) model. 
 
The key issue which 
distinguishes our analysis from the others is that our calculation has
been done in the limit when the average bosonic filling per site is very
large. In this limit, the BH model maps onto a JJA model. A lot of current
calculations on the JJA model are strongly influenced by the results
available for the BH model at low fillings. However, as we stressed in our
previous paper, the algebraic properties of the BH
model in these two limits are very different: in the limit of low fillings,
the model is very close to that of the $SU(2)$ spin model. On the other hand, in the 
limit of large fillings, it is close to that of the $E_{2}$ spin model.

It has
been recently pointed out that $SU(2)$ and the   $E_{2}$ algebras are actually connected by a {\em
singular} transformation\cite{luigi} so that the results obtained in one limit cannot be continued 
adiabatically to those in the other limit, i.e. without crossing a phase boundary. This also lends 
extra support 
to our argument above that the properties of the spin liquids supported by the two spin models are 
distinctly different.
Hence, it is very important to maintain this distinction between $SU(2)$ and $E_{2}$ while performing
calculations on the JJA model. The limit of  large fillings is good for the SC films,
whereas low filling limit is good for helium films\cite{thesis}.

As a second point, a lot of the calculations on JJA model have been done for the
case when an external gate voltage $V_{g}$ is present. We have worked on 
the case $V_{g} = 0$. This situation is close to that of the real
granular SC films. And, finally, our calculations have been done in the 
presence of nearest neighbour interaction $V_{1}$. The presence of
$V_{1}$ brings about non-trivial effects in the JJA model, a quite well
explored instance of which is the onset of supersolidity when rescaled
gate voltage is close to half integer. More details on these differences
may be found in Ref.\cite{thesis}.

There are two previous calculations on the JJA model for $V_{g} = 0$
and
$V_{1} \neq 0$: mean field calculations and numerical simulations\cite{oldrefs}. The
reasons for non-observation of the BM phase in these are as follows: (a){\em Mean
field}: A BM phase benefits partly both from the kinetic energy and the potential
energy terms. But, in a mean field calculation, as soon as superfluidity is
destroyed, the kinetic energy term is completely suppressed. (b){\em Numerical
Simulations}: The problem with this calculation is that the authors investigated only
the superfluid phase in the $V_{0}-V_{1}$ plane at $V_{g} = 0$. Only superfluid
density and structure factor
$S_{\pi,\pi}$ were measured here, which do not distinguish between a
Mott insulator phase and a BM phase: both of these quantities are
zero in these states. More simulations addressing this issue might be helpful
to crosscheck the existence of the BM phase and its properties.

\acknowledgments

We would like to thank Stuart Trugman, Steve Kivelson and Eugene Demler for 
helpful discussions and Nadya Mason for providing us with the MoGe data. Partial support from the
Department of Energy through the complex materials program at the Stanford Synchrotron
Radiation Laboratory is appreciated.

\appendix
\section{World line picture for bosons and fermions}

Here we enlist the actions for bosons and fermions in the world line picture, which might
help the reader in understanding the world line picture discussion of the gauge field
fluctuations in section II.

First, consider a system of $N$ interacting bosons in the world line picture,
{\em without} any gauge field. Partition function of this system is
\[ Z_{B} = \frac{1}{N!}\sum_{P}\int_{\{r_{i}(\beta) = r_{Pi}(0)\}}
        \prod_{i}Dr_{i}(\tau) \exp\left(-\int_{0}^{\beta}d\tau\left[
        \sum_{i}\frac{m}{2}\dot{r}_{i}^{2} + \frac{1}{2}\sum_{i\neq j}
        v(r_{i} - r_{j})\right] \right) \]
where $P$ is the permutation of the particles\cite{negele}.
At low temperatures, the world lines get entangled because of quantum zero
point motion effects, which implies a finite superfluid density for the
bosons\cite{ceperley}.

Now consider an equivalent system of fermions\cite{ceperley}. Here,
the partition function is
\[ Z_{F} = \frac{1}{N!}\sum_{P}e^{i\pi P}\int_{\{r_{i}(\beta)=r_{Pi}(0)\}}
        \prod_{i}Dr_{i}(\tau) \exp\left(-\int_{0}^{\beta}d\tau\left[
        \sum_{i}\frac{m}{2}\dot{r}_{i}^{2} + \frac{1}{2}\sum_{i\neq j}
        v(r_{i} - r_{j})\right] \right) \]
the extra phase factor coming from the anticommutativity of the fermions.
Because of this phase factor, it is readily clear that entangled
configurations enter with random signs and cause cancellation of several
terms. Thus, since the entangled configurations enter the partition function
with low weight, this means that these
configurations are high energy configurations and disentanglement is
favoured for a fermionic system. And, hence the ground state of a Fermi
system is a non-superfluid.

The case of bosons with gauge field fluctuations discussed in section II is intermediate
between these two cases.

\newpage
\begin{figure}
\centerline{\psfig{figure=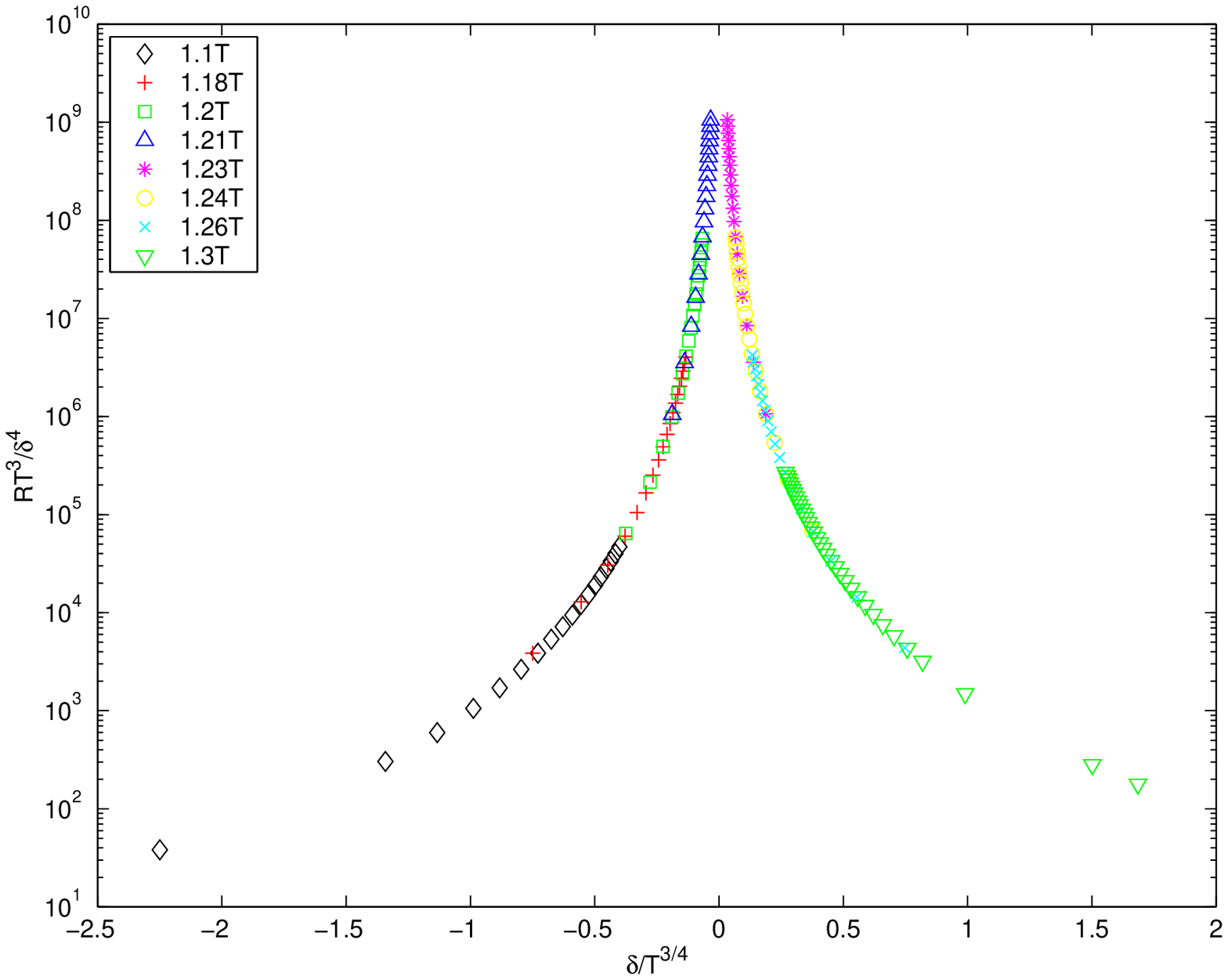,width=16cm}}
\caption{Scaling collpase of the MoGe data from Ref.2 across the metal insulator
transition on a log-linear plot using eqn.(5): $\delta = H - H_{c}$; $H_{c} = 1.22\textup{T}$ is
the critical field at which the metal insulator transition occurs. The data shown covers the entire 
low temperature range of measurement: $\sim 0.02-0.2\textup{K}$.}
\label{fig1}
\end{figure}

\end{document}